\documentclass{article}
\usepackage{spconf,amsmath,amssymb, amsfonts,hyperref, graphicx}
\usepackage[noadjust]{cite}
\usepackage{xcolor}
\usepackage{subcaption}
\usepackage{booktabs}
\usepackage{multirow}
\usepackage{adjustbox}
\usepackage[most]{tcolorbox}
\usepackage{xcolor}
\usepackage[inline]{enumitem}
\definecolor{mypink}{HTML}{A167A5}

\newcommand{\topk}{\texttt{TopK~}}

\title{Sparse Autoencoders Make Audio Foundation Models more Explainable}
\name{\shortstack{Théo Mariotte$^1$ \qquad Martin Lebourdais$^1$ \qquad Antonio Almudévar$^2$\\ \qquad \textit{Marie Tahon}$^1$ \qquad \textit{Alfonso Ortega}$^2$ \qquad \textit{Nicolas Dugué}$^1$}}
\address{$^{1}$ LIUM, Le Mans Université, $^{2}$ VivoLab, I3A, University of Zaragoza.\thanks{This work was funded by PULSAR regional grant 182822 and was performed using HPC resources from GENCI–IDRIS (Grant 2025-AD011016588).}}

\begin{document}
\ninept
\maketitle
\begin{abstract}
Audio pretrained models are widely employed to solve various tasks in speech processing, sound event detection, or music information retrieval.
However, the representations learned by these models are unclear, and their analysis mainly restricts to linear probing of the hidden representations.
In this work, we explore the use of Sparse Autoencoders (SAEs) to analyze the hidden representations of pretrained models, focusing on a case study in singing technique classification. We first demonstrate that SAEs retain both information about the original representations and class labels, enabling their internal structure to provide insights into self-supervised learning systems. Furthermore, we show that SAEs enhance the disentanglement of vocal attributes, establishing them as an effective tool for identifying the underlying factors encoded in the representations.
\end{abstract}

\section{Introduction}
\label{sec:intro}

There is currently an ongoing effort of the scientific community to explainable AI (XAI). 
One line of research focuses on elucidating the decision-making process of AI systems by examining the roles of individual components, such as feed-forward network layers~\cite{geva2020transformer} or self-attention mechanisms~\cite{wiegreffe-pinter-2019-attention}, or by proposing inherently interpretable architectures like induction heads~\cite{olsson2022context}. Our work is more closely aligned with a complementary line of research that seeks to explain complex architectures by interpreting the representations they construct from data. Significant progress has been made in this direction for text and images, with the literature offering methods to analyze word embeddings~\cite{murphy2012learning} and to interpret neural activations in large language models with respect to the input sequences that trigger them~\cite{cunningham_sparse_2023}.
However, research on interpreting the representations of audio processing models remains relatively scarce~\cite{akman2024audio,parekh2024tackling,paissan24a}.

Recently, audio research focused on the development of self-supervised learning models (SSL), especially the so-called \textit{audio foundation models}, which can be used in a wide variety of downstream tasks, such as speech processing, sound event monitoring, or  music understanding.
In speech processing, Wav2vec 2.0~\cite{baevski2020wav2vec} paved the way for models such as HuBERT~\cite{hsu2021hubert} and WavLM~\cite{chen2022wavlm}, which achieve state-of-the-art performance across benchmarks~\cite{yang2021superb, zaiem2023speech}. In audio scene analysis, pretrained models, largely based on the Audio Spectrogram Transformer (AST), have shown strong results in audio tagging and sound event detection~\cite{gong2021ast,chen2022beats,koutini2021efficient}. For music, foundation models such as MERT~\cite{li2023mert} and MuQ~\cite{zhu2025muq} extend self-supervised learning to tasks ranging from pitch and singing technique detection to genre recognition~\cite{zhou2025layer}.
While pretrained models achieve strong performance across audio tasks, their learned acoustic structures remain poorly understood. Recent work has begun addressing interpretability by probing latent representations in domains such as speech~\cite{pasad2023comparative}, music~\cite{zhou2025layer}, and bioacoustics~\cite{cauzinille24_interspeech}. 
Here, we aim to advance this line of research by interpreting the acoustic mechanisms of audio foundation models and identifying minimal components underlying distinct acoustic processing and decision making.
In order to get minimal components, we need to extract fine-grained structure from latent representations, with a high level of sparsity so that few dimensions represent a given factor~\cite{murphy2012learning}.

Sparse autoencoders (SAEs) are good candidates for explaining deep representations, as they project the latent representation in a high-dimensional space with sparsity constraints.
The data are therefore encoded using only a few dimensions, each of which is activated by a limited subset of the data, making these dimensions resemble interpretable features.
Sparsity can be achieved using various techniques, such as the L1-norm constraint~\cite{cunningham_sparse_2023}, TopK L0 constraint~\cite{gao2024scaling}, or BatchTopK~\cite{bussmann2024batchtopk}.
SAEs have been widely used in Natural Language Processing (NLP)~\cite{cunningham_sparse_2023, jing_sparse_2025} and computer vision~\cite{fel_archetypal_2025} to link activations with words or visual patterns.
However, their application to audio models remains limited. 
In~\cite{singh2025discoveringsteeringinterpretableconcepts}, authors apply \topk SAEs to music generative model hidden representations to discover concepts. 

In this paper, we apply \topk SAEs to four audio foundation models spanning speech, music, and sound events, using singing technique recognition as the target task. 
After layer selection, SAEs are trained at varying sparsity levels and evaluated for task information encoding and disentanglement of voice attributes. 
Our contributions are threefold: \textit{(i)} as far as we know, the first application of \topk SAEs to SSL audio models, \textit{(ii)} evidence that SAE representations remain effective for audio SSL models, and \textit{(iii)} demonstration that SAEs enhance disentanglement in terms of completeness. 
Code is available at \url{https://github.com/theomariotte/sae_audio_ssl}.

\begin{figure*}[t]
    \centering
    \includegraphics[width=\textwidth]{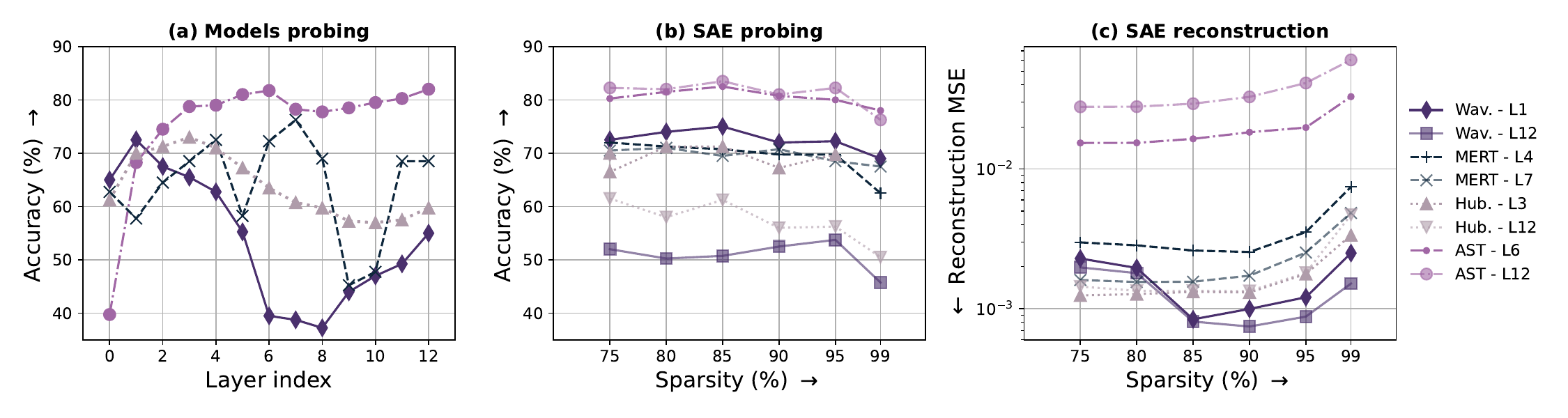}
    \caption{\textbf{(left) Models probing} accuracy on VocalSet test subset using linear probing on the representations from each layer of each model. \textbf{(middle) SAE Probing} accuracy vs. sparsity on the test subset of VocalSet using linear probing on SAE representations from each selected layer of each model. \textbf{(right) SAE Reconstruction} Average MSE vs. sparsity of each SAE in each selected layer on the test subset of VocalSet. }
    \label{fig:sae_eval}
\end{figure*}

\section{Pretrained models and evaluation dataset}

This section presents the pretrained models studied in this paper and the dataset used for our experiments. 
\vspace{-10pt}
\subsection{Pretrained models}

We evaluate four pretrained models. AST~\cite{gong2021ast}, inspired by the Vision Transformer~\cite{dosovitskiy2020image}, tokenizes spectrogram patches for audio tagging; we use the AudioSet-pretrained version\footnote{https://huggingface.co/MIT/ast-finetuned-audioset-10-10-0.4593}~\cite{audioset}. HuBERT~\cite{hsu2021hubert} extends Wav2vec 2.0~\cite{baevski2020wav2vec} by predicting discrete acoustic units from clustered MFCCs; we use the \textit{base} model\footnote{https://huggingface.co/facebook/hubert-base-ls960}. HuBERT was shown to encode low-level (e.g., pitch) acoustic features in early layers and phonetic content in deeper ones~\cite{pasad2023comparative}. WavLM~\cite{chen2022wavlm} follows the same principle as HuBERT but adds overlapped speech augmentation and larger pretraining; we use the \textit{base-plus} model\footnote{https://huggingface.co/microsoft/wavlm-base-plus}, which shows a similar hierarchical pattern as HuBERT for audio features encoding across layers~\cite{pasad2023comparative}. Finally, MERT~\cite{li2023mert}, designed for music understanding, replaces clustering with Residual Vector Quantization autoencoders, achieving state-of-the-art results on Music Information Retrieval (MIR) tasks; we use the \textit{95M} parameter version\footnote{https://huggingface.co/m-a-p/MERT-v1-95M}.  
Each model has 13 layers of transformer, and a hidden vector of $D=768$ dimensions.

\subsection{VocalSet}

We use VocalSet~\cite{wilkins_vocalset_2018}, a 10-hour dataset of 20 professional singers performing various singing exercises (e.g., scales, arpeggios) using various vocal techniques (e.g., vibrato, trill) across vowels. 
Singing technique classification is considered as a target task to analyze sparse representatoins provided by SAEs in the pretrained models.

The choice of VocalSet follows the two following motivations:
\textit{(i)} its manageable size—10 hours of recordings from 20 professional singers across 10 classes, covering exercises such as scales and arpeggios with techniques like vibrato and trill—which facilitates analysis, and \textit{(ii)} its domain relevance—the singing voice shares many acoustic attributes with speech while remaining out-of-domain for most models, enabling effective probing of representations with standard voice descriptors.


\begin{figure*}[t]

\centering
\begin{subfigure}{0.49\textwidth}
    \includegraphics[width=\textwidth]{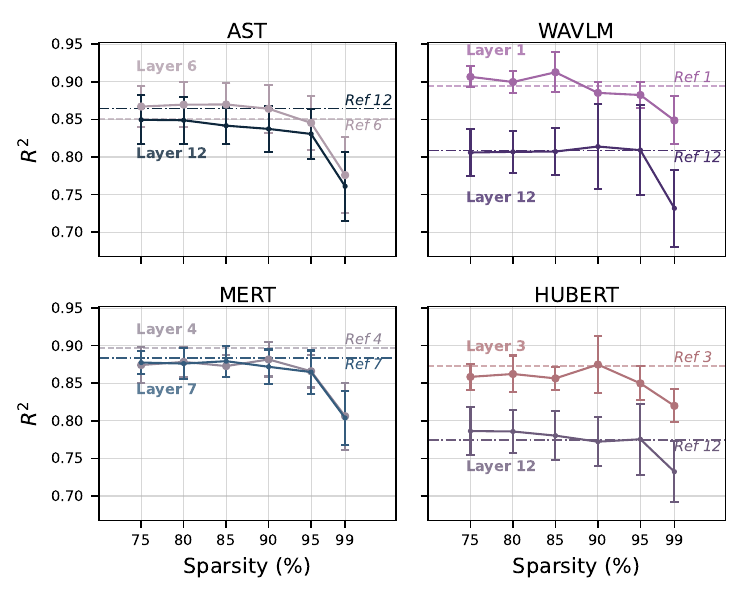}
    \label{fig:first}
\end{subfigure}
\hfill
\begin{subfigure}{0.49\textwidth}
    \includegraphics[width=\textwidth]{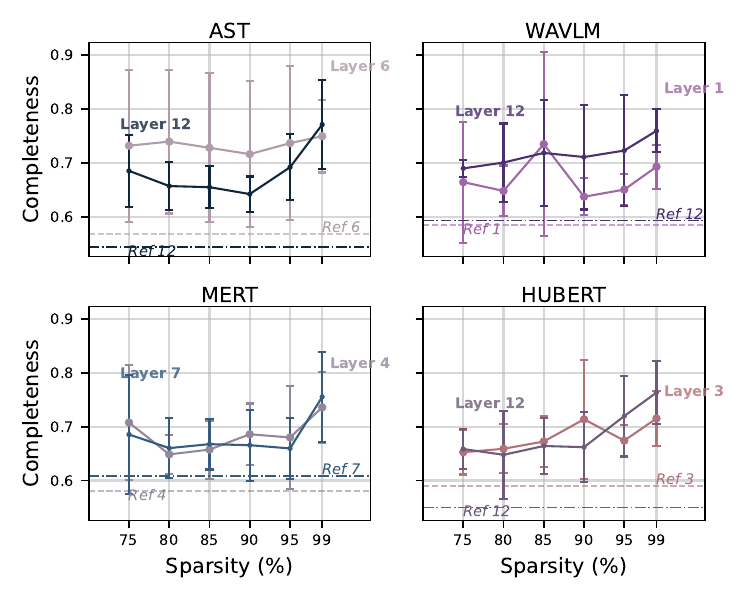}
    \label{fig:second}
\end{subfigure}       
\vspace{-18pt}
\caption{\textbf{(left) Mean and standard deviation of the $R^2$ informativeness of the 10 factors maximizing this metric} vs. sparsity for each model and each layer under study. \textbf{(right) Mean and standard deviation of the completeness of the 10 factors maximizing this metric} vs. sparsity with the same models. The dashed lines represent the value obtained with the original -- non-sparse -- representation.}

\label{fig:r2_comp_eval}
\end{figure*}

\section{SAE Training and Evaluation} \label{sec:sae_training}

Sparse Autoencoders (SAEs) are employed to project hidden representations of SSL models into sparser codes. 
This is expected to improve explainability by mapping dense hidden states into sparse codes where each dimension is more easily attributable to a specific factor of variation. 
This section details the layer selection process, the SAE formulation and training procedure, and their evaluation both in terms of reconstruction quality and downstream performance.  
Unless otherwise stated, models are trained with ADAM optimizer, learning rate of $10^{-3}$, and a batch size of 32.

\subsection{Layer Selection via Linear Probing}
\label{sec:linear_probing}

To identify the layers where information for the task of singing technique classification is most linearly accessible, we apply a linear probing analysis. For each layer of each model, a linear probe is trained on the mean-pooled layer representations. Training is performed on the VocalSet training split with cross-entropy loss. We hold out 20\% of the data for validation and retain the probe with the best accuracy.  

The average probe accuracy across models is shown in Figure~\ref{fig:sae_eval}(left), with the scores of the selected layers summarized in Table~\ref{tab:acc_comp}.
AST achieves the highest accuracy at layers 6 (81.8\%) and 12 (82\%), with stable performance from layers 3 to 12, suggesting that relevant information is equally present and accessible across these layers. WavLM and HuBERT both peak in the early layers (layer 1 (72.5\%) and 3 (73\%), respectively). 
Since these models are known to encode information hierarchically~\cite{pasad2023comparative,cauzinille24_interspeech}, we also include their 12th layer for analysis as a control in the factor prediction experiments (Sect.~\ref{sec:disentanglement}). In contrast, MERT differs from the others, with best accuracy at layers 4 (72.5\%) and 7 (76.2\%).  

Table~\ref{tab:acc_comp} compares the accuracy of probes from the selected layers with results from the literature. 
Probes' accuracy is on par with the literature, with a slight drop in performance with WavLM and HuBERT, which contain no singing voice in their pretraining protocol.


\begin{table}[t]
    \centering
    \caption{\textbf{Vocal technique classification evaluation} on the VocalSet test set. Comparison between linear probes and SOTA performance.}
    \label{tab:acc_comp}
    \begin{adjustbox}{max width=0.8\linewidth}
    \begin{tabular}{llcccc}
   \toprule
    &\textbf{Model} & \textbf{Layer} & \textbf{Acc} (\%) & \textbf{Layer} & \textbf{Acc} (\%) \\
    \cmidrule(r){1-2}
    \cmidrule(lr){3-4}
    \cmidrule(lr){5-6}
    \multirow{3}{*}{\rotatebox[origin=c]{90}{SOTA}}
    & CNN~\cite{wilkins_vocalset_2018} & -  & 80.1 & - & -\\
    & MusicFM~\cite{zhou2025layer} & 5 & 78.3 & - & -\\
    & MuQ~\cite{zhou2025layer} & 6 & 81.5 & - & - \\
    \midrule
    \multirow{4}{*}{\rotatebox[origin=c]{90}{Probes}}
    & AST & 6 & 81.8 & 12 & 82.0 \\
    & WavLM & 1 & 72.5 & 12 & 55.0\\
    & HuBERT &  3 & 73.0 & 12 & 59.8\\
    & MERT & 4 & 72.5 & 7 & 76.2\\
    \bottomrule
    \end{tabular}
    \end{adjustbox}
\end{table}

\vspace{-5pt}
\subsection{SAE Formulation and Training}
\label{sec:sae_formulation}

After identifying the most informative layers for the target task, the next step is to train SAEs on the corresponding representations.
Let $\mathbf{x}_{l} \in\mathbb{R}^{T\times D}$ be the hidden representation of layer $l$, where $T$ is the number of frames (tokens in AST), and $D$ is the dimensionality of the representation.
Since our focus is not on explaining the temporal structure, we apply average pooling along the temporal axis to obtain a fixed-size vector $\bar{\mathbf{x}}_{l} \in \mathbb{R}^D$.

This vector is encoded into a sparse code $\mathbf{z}_l \in \mathbb{R}^N$ ($N > D$) using a \topk SAE~\cite{gao2024scaling}:  
\begin{equation}
    \mathbf{z}_l = \mathrm{TopK}\left(\mathrm{ReLU}\left(\mathbf{W}_{l}^{e} \bar{\mathbf{x}}_l + \mathbf{b}_l^e\right)\right),
\end{equation}
where $\mathbf{W}_{l}^{e}$ and $\mathbf{b}_l^e$ are the encoder trainable parameters. The operator $\mathrm{TopK}(\cdot)$ retains only the $k$ largest activations, with $k$ acting as a sparsity-control hyperparameter. The sparse code is then decoded with a linear decoder (without bias) as $\hat{\bar{\mathbf{x}}}_l = \mathbf{W}_l^d \mathbf{z}_l$.  

The SAE is trained with a mean squared error objective:  
\[
\ell_{MSE}(\hat{\bar{\mathbf{x}}}_l, \bar{\mathbf{x}}_l) = \|\hat{\bar{\mathbf{x}}}_l - \bar{\mathbf{x}}_l\|_2^2.
\]  

For each selected layer, we train six SAEs with sparsity levels ranging from 75\% to 99\%, where e.g., 95\% sparsity corresponds to $k = \lfloor (1-0.95) \times N \rfloor$. Across all models, we set $D=768$ and $N=2048$, and train on the VocalSet training subset. The model with the best validation MSE is retained.

\subsection{Reconstruction Performance}
\label{sec:sae_recon}

Figure~\ref{fig:sae_eval}(right) reports the reconstruction MSE on the VocalSet evaluation set as a function of sparsity. Reconstruction quality degrades as sparsity increases, consistent with findings in LLMs~\cite{gao2024scaling} and CV models~\cite{fel_archetypal_2025}. The audio encoding also plays a role: AST, which is based on spectrogram patches, exhibits MSE values an order of magnitude higher than convolutional encoders applied directly to the waveform. Interestingly, WavLM shows a trade-off, achieving its best MSE at 85\% sparsity.

\subsection{Downstream Performance from Sparse Codes}
\label{sec:lin_prob_sae}

While reconstruction quality is informative, MSE alone does not directly reveal whether task-relevant information is preserved in the sparse codes. We therefore repeat the linear probing protocol on the SAE representations $\mathbf{z}_{l}$ at each sparsity level. Training follows the same setup as in Section~\ref{sec:linear_probing}, and the best validation probe is selected.  

The results, shown in Figure~\ref{fig:sae_eval} (middle), indicate that sparsity has little effect on singing technique classification accuracy compared to the original representations (Table~\ref{tab:acc_comp}, Fig.~\ref{fig:sae_eval} (left)). For instance, AST performance remains stable up to 95\% sparsity, and only drops sharply at 99\%, when merely 20 active dimensions remain.  

\begin{center}
\begin{adjustbox}{max width=0.9\linewidth}

\begin{tcolorbox}[colback=mypink!5,colframe=mypink!80,title=Takeaway]
We first select informative layers through linear probing, then train SAEs at varying sparsity levels. While reconstruction quality degrades as sparsity increases, downstream classification accuracy remains robust until very high sparsity, suggesting that sparse codes retain task-relevant information. The next section investigates whether these representations also disentangle voice attributes.  
\end{tcolorbox}
\end{adjustbox}
\end{center}

\vspace{-10pt}
\section{Disentanglement evaluation}
\label{sec:disentanglement}

This section evaluates the SAEs representations under the scope of disentanglement.
We follow two hypotheses in this section:
\begin{enumerate*}[label=(\roman*)]
    \item SAEs representation improves the linear prediction of voice attributes
    \item they improve the disentanglement of the factors in the representation by requiring fewer dimensions to encode them.
\end{enumerate*}

\subsection{Voice attributes as descriptive factors}

Since the target task of our study is singing technique detection, we rely on low-level speech-related features.
Specifically, we used the OpenSMILE\footnote{https://www.audeering.com/research/opensmile/} library to extract the eGeMAPS descriptive features~\cite{eyben_geneva_2016} for each audio file of the test set (400 files). Although originally designed for emotion recognition in speech, we hypothesize that they might also be relevant to characterize singing voice, as both domains mostly rely on prosody.

The features can be analyzed independently, e.g., to compute disentanglement metrics, or grouped into broader families to capture large-scale behaviors. 
In the second scenario, features are grouped into seven families: \texttt{pitch}, \texttt{loudness}, \texttt{formants} (F1,F2,F3), \texttt{MFCC} (coef. 1 to 4), \texttt{rhythm}, \texttt{spectral} (alpha ratio, slope, spectral flux, HammarbergIndex), and \texttt{quality} (jitter, shimmer, HNR).

\subsection{Disentanglement evaluation metrics}

We evaluate attributes disentanglement using the same approach as DCI~\cite{eastwood2018framework}.
Let $\mathbf{f}_i\in\mathbb{R}^{M}$ be the $i$-th factor extracted from each of the $M$ samples of the test set.
We train a Lasso linear regression $g_i$ that maps the set of SAE representation $\mathbf{Z}_l \in \mathbb{R}^{M\times N}$ to the set of factors $\mathbf{f}_i$.
This process is repeated for each of the $F$ factors.

We evaluate disentanglement using two metrics: (i) informativeness, which measures how accurately a factor can be predicted from the SAE space. It is quantified (in this paper) using the regression coefficient of determination ($R^2$). 
(ii) completeness, which reflects the number of dimensions needed to predict a feature. Completeness ranges from 0 to 1, with higher values indicating that only a few dimensions in $\mathbf{z}_l$ are sufficient to predict the factor $\mathbf{f}_i$.



\begin{figure}[t]
    \centering
    \includegraphics[width=0.6\linewidth]{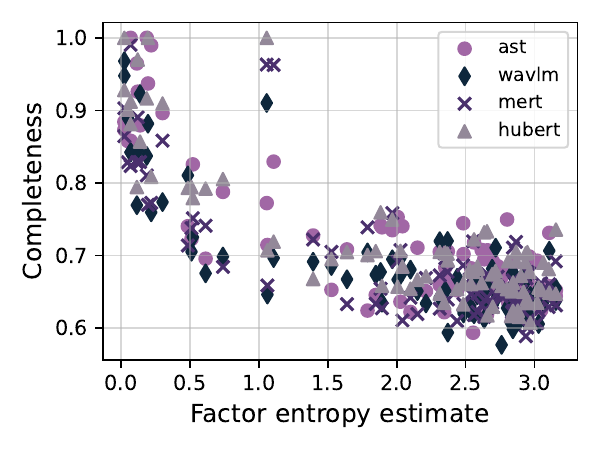}
    \caption{\textbf{Completeness vs. estimated entropy of the factors' distribution} for 88 eGMAPS factors for each model at 90\% sparsity level.}
    \vspace{-15pt}
    \label{fig:comp_vs_entropy}
\end{figure}

\subsection{Results}

\subsubsection{Informativeness and Completeness}

Figure~\ref{fig:r2_comp_eval} shows the informativeness and the completeness of the top-10 factors.
It means that for each metric, we select the 10 factors obtaining the highest informativeness (resp. completeness) score. 
This allows to reduce the set of factors to analyze, which also facilitates the analysis in Section~\ref{sec:factors_id}.

Figure~\ref{fig:r2_comp_eval} (left) presents the $R^2$ as a function of sparsity for each layer in each model under study.
The \textit{ref} dashed line indicates the $R^2$ obtained from the original representation. 
Overall, trends are consistent across models and layers: SAEs generally preserve the informativeness of the representation, with degradation only observed at very high sparsity levels. 
This is in line with the linear probing performed in section~\ref{sec:lin_prob_sae}. 

Figure~\ref{fig:r2_comp_eval} (right) presents the completeness score as a function of sparsity for each model. 
Contrary to informativeness, completeness increases with sparsity. 
Furthermore, SAEs learn more \textit{complete} representations compared to the original representations. 
This behavior is expected: when informativeness remains constant, but fewer active dimensions are available, the representation must concentrate the predictive information into fewer units, thereby increasing completeness.

Finally, Figure~\ref{fig:comp_vs_entropy} presents the Completeness as a function of the factors' entropy. 
The entropy is estimated using $k$-Nearest Neighbor estimator~\cite{kozachenko1987sample}.
The results show that factors with higher entropy exhibit lower completeness, which is consistent with Shannon’s information theory: variables with higher entropy require more bits to be encoded.

\vspace{-5pt}
\subsubsection{Factors identification}
\label{sec:factors_id}
\begin{figure}[t]
    \centering
    \includegraphics[width=\linewidth]{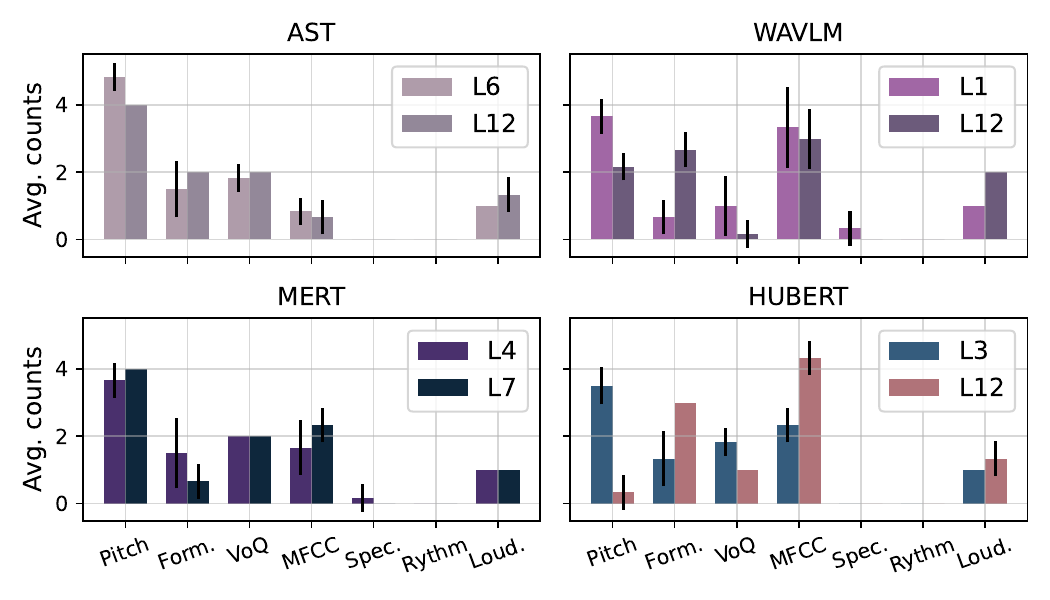}
    \caption{\textbf{Count of the 10 best-predicted ($R^2$) factors} averaged across sparsity for each layer of each model under study. Factors are grouped by families to simplify the visualization.}
    \vspace{-15pt}
    \label{fig:best_r2_feat}
\end{figure}

In this last section, we explore the distribution of the best predicted factors. 
Figure~\ref{fig:best_r2_feat} shows the average occurrence of each family of factor, with average and standard deviation computed across sparsity levels. 
The low error bars show that sparsity has little impact on the type of feature that can be reconstructed.
However, the layer index has a dominant impact, mostly for WavLM and HuBERT. 
For example, pitch information is accurately predicted by the first layers (1 and 3, respectively) while formants information emerges in the final layer (12). 
This follows the literature~\cite{pasad2023comparative} showing that phoneme-related information is located in the last layers.

\begin{center}
\begin{adjustbox}{max width=0.9\linewidth}

\begin{tcolorbox}[colback=mypink!5,colframe=mypink!80,title=Takeaway]
SAEs preserve the informativeness of SSL representations while significantly increasing completeness, 
i.e., fewer dimensions are needed to predict generative factors. 
Completeness decreases with factor entropy, as expected from information theory, and factor prediction patterns reveal that 
pitch information emerges in early layers, whereas formant and phoneme-related information concentrate in higher layers.
\end{tcolorbox}
\end{adjustbox}
\end{center}

\section{Conclusions}
\label{sec:ccl}

In this paper, we propose a methodology to explain the data representation of audio foundation models.
To achieve this, we explore the use of \topk sparse autoencoders (SAEs), applied to four different pretrained models: AST, WavLM, HuBERT, and MERT.
Experiments are conducted within the task of singing technique recognition. 
After identifying the most informative layers for this task, we train SAEs at different sparsity levels. 
A first analysis shows that sparse representations retain task information, even at extreme sparsity levels.
In a second analysis, we explore capacities of SAEs to disentangle acoustic factors. 
While factors' prediction (informativeness) is not improved by sparse representations, results show that SAEs require less dimensions to encode the same information, i.e., better completeness.
Finally, we show that the best reconstructed factors can be identified, and we highlight a relationship between factors' entropy and completeness.

\bibliographystyle{abbrv}
\bibliography{refs}

\end{document}